\documentclass{article}
\usepackage{spconf,amsmath,graphicx}

\usepackage{microtype}
\usepackage{paralist}
\usepackage{url}
\usepackage{hyperref}
\usepackage{nicefrac}

\usepackage{booktabs}
\usepackage{multirow}
\usepackage{siunitx}
\usepackage{caption}
\usepackage{subcaption}

\ninept

\usepackage{color}
\newcommand{\alex}[1]{\textcolor{blue}{#1}}

\newcommand{\pavan}[1]{\textcolor{magenta}{#1}}

\renewcommand{\alex}[1]{\textcolor{black}{#1}}

\renewcommand{\pavan}[1]{\textcolor{black}{#1}}

\title{ASPED: An Audio Dataset for Detecting Pedestrians}
\name{Pavan Seshadri\textsuperscript{b}, Chaeyeon Han\textsuperscript{a}, Bon-Woo Koo\textsuperscript{d}, Noah Posner\textsuperscript{c}, Subhrajit Guhathakurta\textsuperscript{a}, Alexander Lerch\textsuperscript{b}\thanks{This work was funded by NSF Award 2203408.}}
\address{\textsuperscript{a}CSPAV, \textsuperscript{b}Music Informatics Group, and \textsuperscript{c}IPaT,  Georgia Institute of Technology\\ \textsuperscript{d}Toronto Metropolitan University}
\begin{document}
%
\maketitle
\begin{abstract}
We introduce the new audio analysis task of pedestrian detection and present a new large-scale dataset for this task. While the preliminary results prove the viability of using audio approaches for pedestrian detection, they also show that this challenging task cannot be easily solved with standard approaches.
\end{abstract}
\section{Introduction}
\label{sec:intro}
    The intelligent analysis of urban soundscapes plays an increasingly important role in the design of smart cities. Microphones can complement or even replace other forms of sensors because
    \begin{inparaenum}[(i)]
        \item   they are affordable, 
        \item   have low power requirements, 
        \item   can cover large angles up to 360 degrees, and 
        \item   are not negatively impacted by light conditions, weather patterns such as fog, or obstacles blocking the angle of view.
    \end{inparaenum}
    
    
    In this paper, we propose a new challenging task in urban sound analysis: the detection of pedestrians from audio-only signals.
    The detection of pedestrians helps in alleviating bottlenecks and in triggering advance warnings about potential dislocations. Understanding the temporal and spatial variation in demand for pedestrian infrastructure can also lead to better resource use, more equitable service delivery, and greater sustainability and resilience. 
    Detecting pedestrians through audio poses some unique challenges. The audio signal pedestrians produce are often low volume and can be as diverse as steps and speech. These signals have to be detected in a poly-timbral and time-varying mixture of multiple urban sound sources with overlapping frequency content.
    
    To allow investigation of the viability of this novel task as well as to enable and encourage future research on the task of pedestrian detection through audio signals, we present a new, large-scale dataset containing audio and video data recorded in multiple separate recording sessions at different locations at the Georgia Tech campus, Atlanta.\footnote{\href{https://urbanaudiosensing.github.io/ASPED}{urbanaudiosensing.github.io/ASPED}, last access date Sep 6, 2023} The number of pedestrians in proximity to the microphones is annotated through video analysis with three different proximity radii. 

    The main contributions of this paper are
    \begin{inparaenum}[(i)]
        \item   the introduction of a new task in urban sound analysis: pedestrian detection,
        \item   the publication of a new large-scale audio dataset for this task called ASPED (Audio Sensing for PEdestrian Detection), and
        \item   the presentation of baseline results for benchmarking and for viability analysis.
    \end{inparaenum}

\section{Related work}
\label{sec:relwork}
Identifying the environmental context through Sound Event Detection (SED) has been an active area of research in the past decade \cite{babaee_overview_2017, mesaros_sound_2021}. The challenge of SED in a typical outdoor environment is the detection of an event from multiple known and unknown sources of sound that are emitted simultaneously. Initial approaches to SED have used Mel-frequency cepstral coefficients (MFCC) or other time-frequency representations such as Fourier transform and the wavelet transform \cite{parascandolo_recurrent_2016, mesaros_acoustic_2010}. Other approaches included non-negative matrix transformations (NMF) and spectrogram analysis with image processing techniques \cite{innami_nmf-based_2012, dessein_real-time_2013, dennis_overlapping_2013}. Recent advances in feedforward neural networks (FNN) and multilabel recurrent neural networks (RNN) have been particularly promising for SED \cite{cakir_convolutional_2017, cakir_polyphonic_2015, parascandolo_recurrent_2016}. 

The advances in SED have led to a small but emerging field focusing on the detection and classification of urban sounds \cite{babaee_overview_2017, mesaros_sound_2021}. This research has been instrumental in the automatic detection of crime indicators such as screams and gunshots and in monitoring urban noise pollution \cite{7098998, valenzise_scream_2007, hammad_unsupervised_2023, socoro_anomalous_2017}. A large-scale research effort in this domain has been an NSF-funded project called SONYC for detecting noise and tagging urban sound sources \cite{cartwright_sonyc_2019}. This project has provided a large dataset of audio recordings tagged by citizen science volunteers who annotated the presence of 23 fine-grained categories of events. Another such dataset is AudioSet, which was developed by the Machine Perception Research Organization at Google \cite{gemmeke_audio_2017}. AudioSet is a large-scale collection of human-labeled \SI{10}{\second} sound clips from over 2 million YouTube videos and contain 527 classes of annotated sounds. The same group at Google has also released the YouTube-100M data set labeled with one or more topic identifiers from a set of 30,871 labels \cite{hershey_cnn_2017}. These labels are assigned automatically based on the metadata and image content. A number of labeled data sets for SED have also been developed from contributions to freesound.org, including ESC-50 and FSD50K \cite{piczak_esc_2015, fonseca_fsd50k_2022}. In addition, VGGSound is another audio-visual dataset released in 2020 containing more than 310 audio classes \cite{chen_vggsound_2020}.  However, previous research have not focused on sensing pedestrians using SED techniques. Thus, existing datasets are not annotated with pedestrian counts and cannot be easily relabeled. Furthermore, capturing meaningful audio data from pedestrians is particularly difficult because it is not clear what kind of sound can clearly identify pedestrian movement -- is it footsteps, conversation, rustling of clothes, or a combination of these components? The challenge also is to detect such sound in a loud multilayered soundscape within a street environment with vehicular noise. 

\section{Dataset}
\label{sec:dataset}

We aimed to reach the following goals by planning and creating the dataset:
\begin{inparaenum}[(i)]
    \item large scale: large amount of data to accommodate advanced data-hungry machine learning models,
    \item high quality: to allow for investigation of the impact of audio quality levels (sample rate, word length), recordings should be provided in high audio quality, and
    \item diversity: different recording locations and times to account for a variety of scenarios.
\end{inparaenum}
\begin{figure}
    \centering
    \includegraphics[trim={0 10cm 0 32cm},clip,width=\columnwidth]{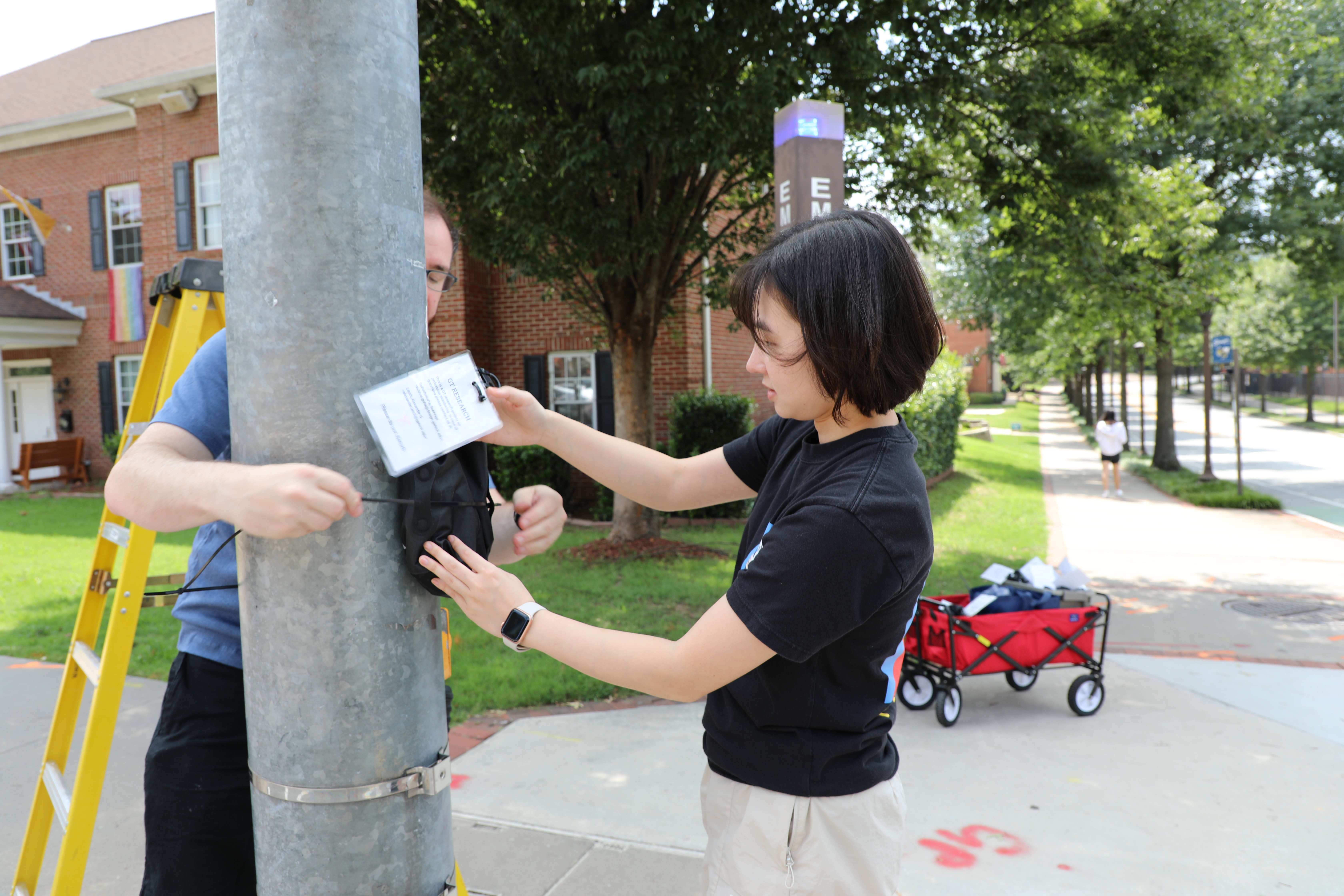}
    \caption{Research team installing audio recorders in the field.}
    \label{fig:installaudio}
\end{figure}

\subsection{Data acquisition} 
Two hardware setups were used for data acquisition. The audio collection setup consisted of multiple Tascam DR-05X audio recorders 
with power banks for extended duration recording, Saramonic SR-XM1 microphones, and 5L OverBoard Waterproof Dry Flat Bags for audio permeable weatherproofing. 
The video  setup is a GoPro HERO9 Black cameras 
with power banks (housed in a Seahorse 56 OEM Micro Hard Case) for extended duration recording. 

Multiple audio sensors and cameras were deployed for each data collection session. 
For each session, the recorders were placed in their weatherproof bags  once started, then secured to their recording locations using zip ties. Recorders were secured at approx.\ chest height as it was determined that sub-meter variation in height did not affect audio quality. Figure~\ref{fig:installaudio} shows the installation of one audio recorder. 

The cameras were set to time-lapse mode with a \SI{1}{\second} duration. Wi-Fi functionality was disabled to extend battery life. Multiple cameras were utilized to keep all recorders in view. The camera mounts were secured at $\approx$\SI{2.5}{\meter} using zip ties. 

In order to time sync the cameras, the time as listed on www.time.gov was shown on a mobile device to each camera after starting recording. A fox 40 pearl whistle was then blown and the precise time was recorded. This whistle was used to sync the audio recorders. In deployment locations over larger areas, multiple whistle blows were conducted. 

Recorders were deployed at two on-campus locations, the Cadell Courtyard, and the Tech Walkway. Both locations are near areas with restaurants and cafes but are off-limits to vehicular traffic. The battery life of the recording devices limited the length of each recording session to approx.\ 2 days per session. 

In total, we captured 1-fps video recordings that sum up to 3,406,229 frames and the corresponding audio recordings of nearly 2,600 hours. 
All but one recorded days are weekdays. 




\subsection{Annotations}\label{sec:annot}
The number of pedestrians that actually passed the audio recorders was detected and annotated by applying the Masked-attention Mask Transformer (Mask2Former) \cite{cheng_masked-attention_2021}, with a prediction threshold of 0.7 on video the recordings. This study used a Mask2Former implementation by OpenMMLab,\footnote{\href{https://openmmlab.com}{openmmlab.com}, last access date Sep 5, 2023} trained on Microsoft COCO \cite{lin_microsoft_2014}.

\begin{figure}
    \centering
    \includegraphics[trim={0 .5cm 0 1.5cm},clip,width=\columnwidth]{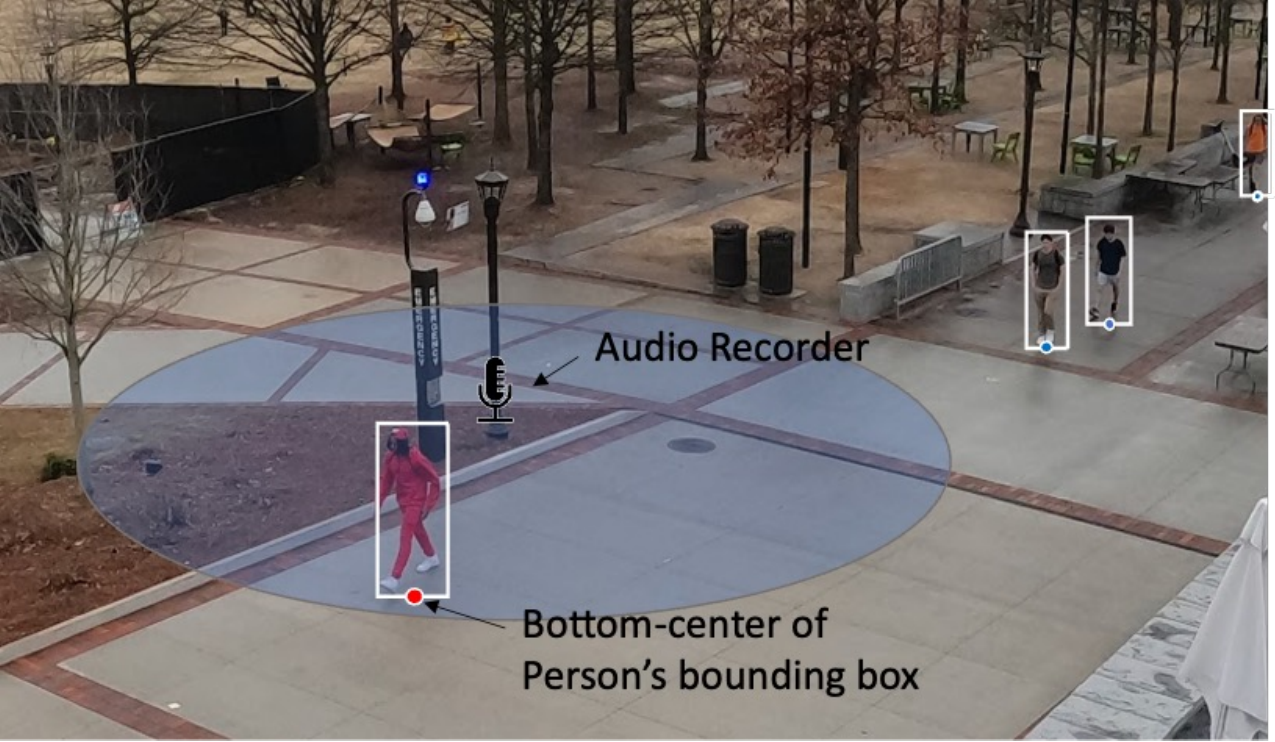}
    \caption{Pedestrian detection video setup.}
    \label{fig:recorders}
\end{figure}
For each video frame, bounding boxes of the detected ‘person’ class were first extracted from the prediction from Mask2Former. Next, circular buffers of different radius $r\in[\SI{1}{\meter},\SI{3}{\meter},\SI{6}{\meter},\SI{9}{\meter}]$ were overlaid on the video frames around the poles to which audio recorders were attached. The buffers were angled to match the perspective of each video recording instance. Finally, the number of pedestrians with the bottom center of the bounding box intersecting with recorder buffers was counted and labeled in each frame. Figure~\ref{fig:recorders} visualizes an example buffer and pedestrians with bounding boxes.


Each frame has four sets of annotation data for the four different recording radii (see Sect.~\ref{sec:annot}). Among the annotated videos, frames without any detected pedestrians were the most common. 
Frames with one pedestrian were next most frequent, followed by those with two, then three, four pedestrians, and so on. Proportions of each count within the labels for each radius are displayed in Table ~\ref{tab:percentages}. The labeled data contains more pedestrians detected during the daytime as shown in Fig.~\ref{fig:data-hist}. Pedestrian activities were at peak around noon, especially during lunch time (11AM--14PM).

\section{Experiments}
\label{sec:results}

\subsection{Experimental setup}
We determine a baseline level of performance for this task with three different models, all targeting a binary classification (pedestrians present/not present) at different microphone radius settings and for different pedestrian count thresholds separating the present/not present classes. \pavan{We aim to investigate the effects of these permutations to establish an understanding of how a basic classifier responds to the change of data parameters of this task.}

\begin{figure}[t]
    \centering
    \includegraphics[width=\columnwidth]{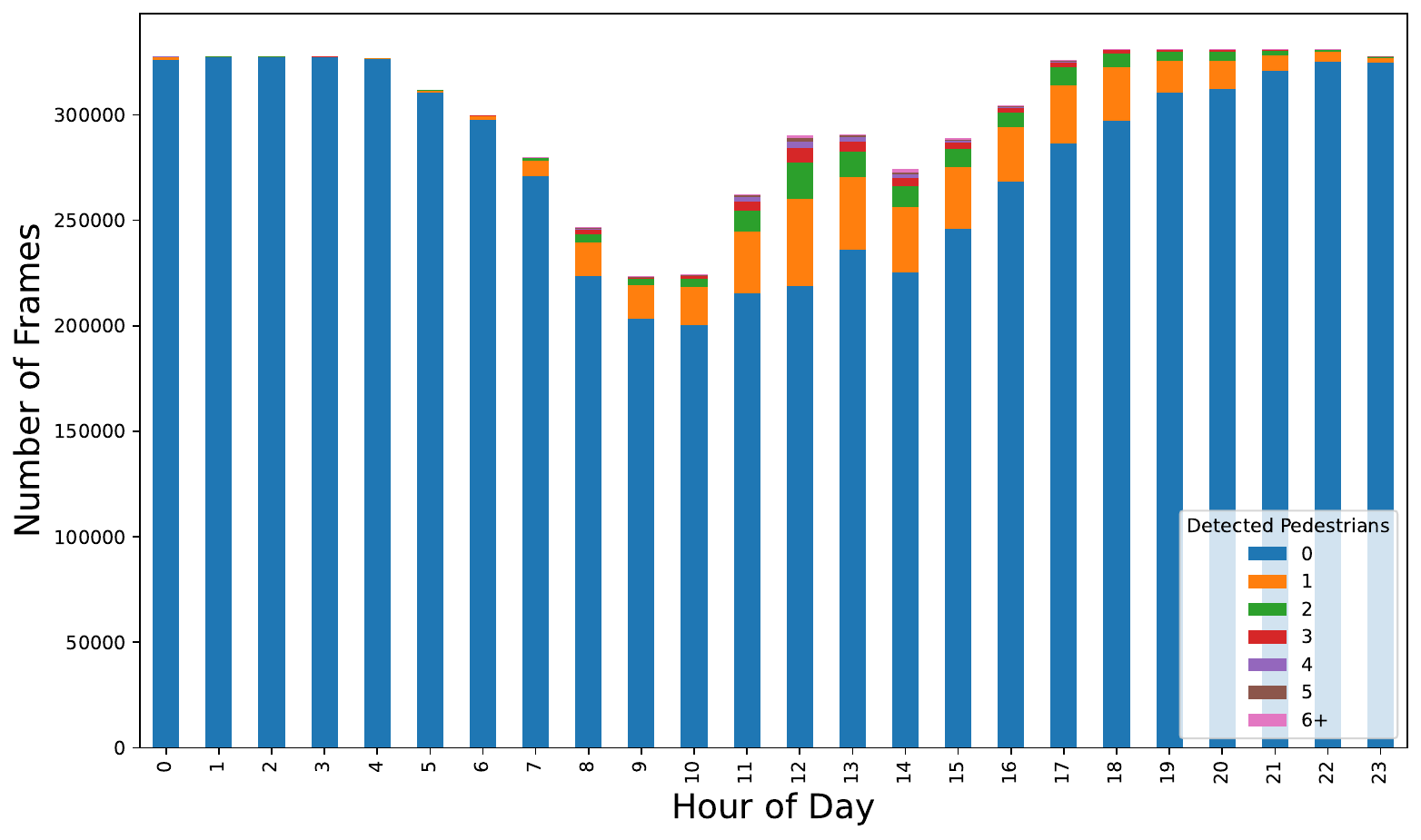}
    \caption{\label{fig:data-hist}Number of pedestrians radius $r=\SI{6}{\meter}$ by hour of day.}
    \label{figuur:opnemer}
\end{figure}

\begin{table}
\centering
\begin{tabular*}{\columnwidth}{c| @{\extracolsep{\fill}}c|c|c|c|c}
 & \multicolumn{4}{c}{\textbf{Pedestrian Counts}} & \\
\hline
\textbf{Radius}& \textbf{0} & \textbf{1} & \textbf{2} & \textbf{3} & \textbf{4+} \\
\hline
1 & 99.4757 & 0.4723 & 0.0453 & 0.0058 & 0.0008 \\
\hline
3 & 97.2538 & 2.1906 & 0.4075 & 0.0962 & 0.0519 \\
\hline
6 & 91.9113 & 5.2566 & 1.7578 & 0.6262 & 0.4481 \\
\hline
9 & 86.6895 & 7.7196 & 3.0175 & 1.2999 & 1.2734 \\
\end{tabular*}
\caption{Percentage proportion of each pedestrian count per the labels for each recorder radius}
\label{tab:percentages}
\end{table}

\subsubsection{Model architectures}
First, we investigate using the  VGGish embeddings \cite{hershey_cnn_2017}, pre-trained on AudioSet \cite{gemmeke_audio_2017}, as input to a transformer encoder to learn temporal relationships across each segment (referred to as \textit{VGGISH}). 
Second, we use a convolutional encoder with a log-mel spectrogram input, followed by the aforementioned transformer encoder (referred to as \textit{CONV}). 
Third, we explore using the Audio Spectrogram Transformer, which has been shown to deliver state-of-the-art performance for audio scene classification tasks \cite{AST} (referred to as \textit{AST}). 
All models compute class output probabilities through an appended linear classification layer with a sigmoid activation function.

\subsubsection{Feature extraction}
All network inputs are extracted in time frames of approx.\ \SI{1}{\second} length. Both VGGISH and CONV follow the pre-processing procedure for the pre-trained VGGish network \cite{hershey_cnn_2017}, resulting in a 128-dimensional VGGish embedding or a $96\times64$ dimensional (time $\times$ freq) log-mel spectrogram, respectively. The AST input is a spectrogram with dimensionality $100\times128$ (time $\times$ freq), following the original publication \cite{AST}.

The input of the VGGish and CONV models are a sequence of 10 concurrent features, corresponding to each \SI{1}{\second} frame per \SI{10}{\second} audio segment. The input to the AST are single features per \SI{1}{\second} frame. Each classification is done per frame, for every second of audio.

\subsubsection{Training procedure}
As our data contains pedestrian counts per frame, we create classification labels where values of 0 are counted as negative-activity, and any value above 0 is counted as positive-activity. 

The dataset was randomly split into train/test/validation subsets with 80/10/10 proportion, respectively.  For testing and validation, any overlapping segments are removed so that labels are not re-used multiple times. 

The loss function for all models is binary cross-entropy. 
As Fig.~{\ref{fig:data-hist}} shows, the label distribution is highly skewed towards no-activity; To promote the learning of pedestrian activity, we use the following augmentations for the underrepresented classes: 
\begin{inparaenum}[(i)]
    \item \textit{weighted batch sampling} --- in each mini batch, audio segments are sampled with replacement such that roughly half will contain at least one pedestrian activity event;
    \item \textit{variable weighted loss} --- each classes loss is weighted dynamically per batch relative to its density in the training samples, such that both positive and negative pedestrian-activity contribute roughly equally to the loss per batch. The weighting function used is shown below: 
    \begin{equation}
        \mathcal{L} = \lambda\mathcal{L}_\mathrm{BCE+} + (1-\lambda)\mathcal{L}_\mathrm{BCE-}
    \end{equation}
    \begin{equation}
    \lambda = \left\{
    \begin{array}{lr}
        \frac{\nicefrac{1}{num^{+}}}{\nicefrac{1}{num^{+}} + \nicefrac{1}{num^{-}}}, & \text{if } num^{+} \neq 0\\
        0, & \text{if } num^{+} = 0
    \end{array}\right.    
    \end{equation}
\end{inparaenum}


\subsubsection{Hyperparameters and implementation}
For CONV and VGGISH, we use 1 transformer encoder with 4 attention heads, with a hidden dimensionality of 128. CONV contains 6 convolutional blocks each containing a conv2D, batchnorm, and leakyReLU layer. Both networks are trained with a learning rate of 0.0005. For the AST, we use the base configuration per the authors implementation\footnote{\href{https://github.com/YuanGongND/ast}{github.com/YuanGongND/ast}, last access date Sep 5, 2023} pre-trained on ImageNet \cite{IMAGENET} and AudioSet \cite{gemmeke_audio_2017} with hidden dimensionality of 768, and a learning rate of 5e-7. We train the CONV and VGGISH models for 20 epochs and the AST for 10 epochs, with the best performing model selected via performance on the validation set. Parameters are optimized using the ADAM optimizer \cite{ADAM}. We use a batch size of 256 for VGGISH and CONV, and 32 for AST. 

     \begin{figure}
         \centering
        \includegraphics[width=\linewidth]{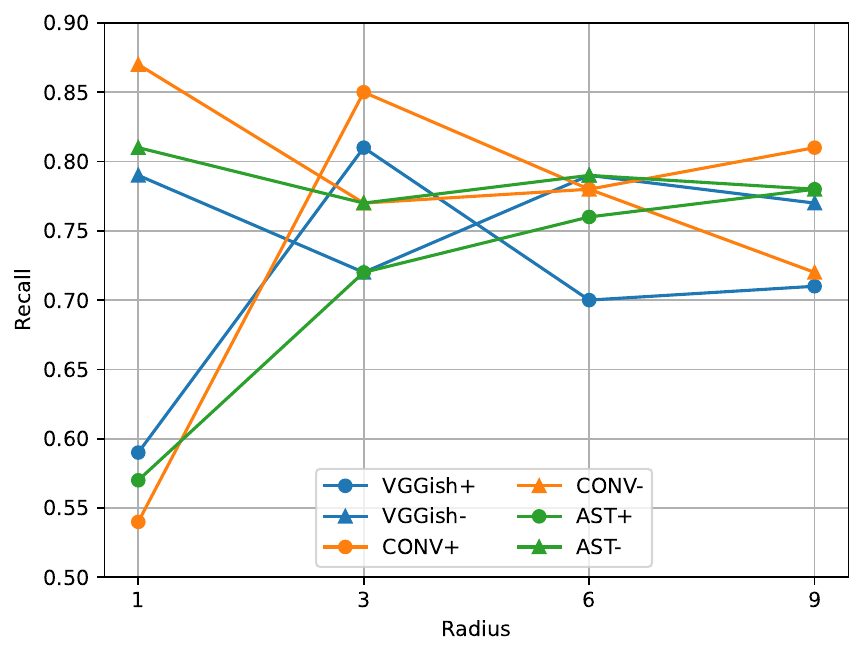}
        \caption{\label{all-res}Recall for each class over recording radius. Positive and negative classes are denoted by "+" and "-", respectively.}
     \end{figure}

\subsubsection{Experiments}
We evaluate the baseline performance measured by class-level and macro-average recall with the following experiments:

\textbf{E1 --- Comparison of baseline architectures:} 
In order to capture task performance using general audio classification methods as well as to evaluate performance across architectures of varying complexity, the three models introduced above are compared. The complexity ranges from $\sim100\mathrm{K}$ trainable parameters  (VGGISH) to $\sim80\mathrm{M}$ trainable parameters  (AST). 

\textbf{E2 --- Impact of recording radius on accuracy:} 
With this experiment, the impact of the recording radius on the performance is investigated. Spatial consideration for determining pedestrian activity affects both the count and diversity of pedestrian noises: smaller radii contain a lower number of pedestrians that should be easier to classify while larger radii contain a higher number of pedestrians with harder to classify samples. As such, larger radii should provide a greater diversity of pedestrian signals to our models with the downside that counted pedestrians are more difficult to detect. 

\textbf{E3 --- Impact of pedestrian count during training and testing on performance:} 
As the threshold for binary classification can be set at arbitrary pedestrian counts, and does not necessarily have to be identical for both training and testing. Therefore, we determine the impact of different training thresholds on different inference thresholds and thus investigate the model generalizability to different pedestrian activity. 
This experiment utilizes a CONV model at radius $r=\SI{6}{\meter}$ while thresholding labels with values $p_\mathrm{T} \in$ [1,2,3,4] such that any value lower than $p_\mathrm{T}$ is set to 0. We then test each trained model on the 4 resulting test sets.
     \begin{figure}
         \centering
            \includegraphics[width=.93\linewidth]{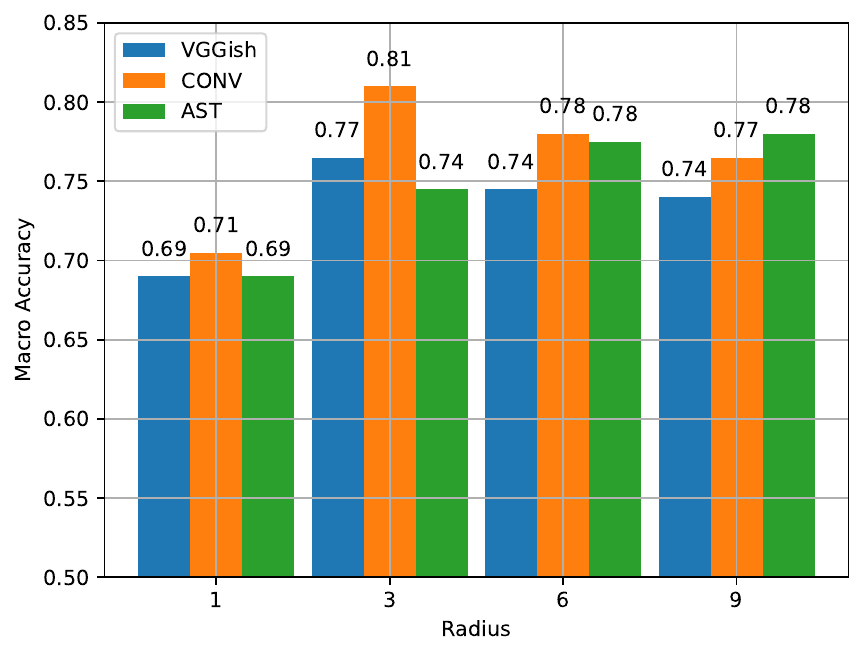}
            \caption{\label{all-avg}Macro average accuracy using the VGGISH, CONV, and AST models.}
     \end{figure}
\subsection{Results}

\textbf{E1:} Figures~\ref{all-res} and \ref{all-avg} detail the results using the VGGISH, CONV, and AST models, respectively. We can make the following observations. First, the VGGISH model is in most cases outperformed by both the CONV and AST models\alex{, which is expected as the pre-trained VGGish embeddings are not fine-tuned for pedestrian detection}. \pavan{Second, in terms of macro accuracy, the performance of the VGGISH and AST model are fairly constant across radii 3 to \SI{9}{\meter}, while the CONV model achieves highest performance on radii 3 and \SI{6}{\meter}.} \alex{The reasons for these radii working better could be a combination of less imbalanced class distribution and reasonable proximity to the microphone.} Third, the AST generally has closest parity between performance on both classes. Lastly, the negative class recall seems to slightly outperform the recall for the positive (pedestrian) activity, although the dramatic class imbalance observed in the data is not reflected in the results showing the effectiveness of the sampling and loss weighting applied during training.

     \begin{figure}
         \centering
        \includegraphics[width=.925\linewidth]{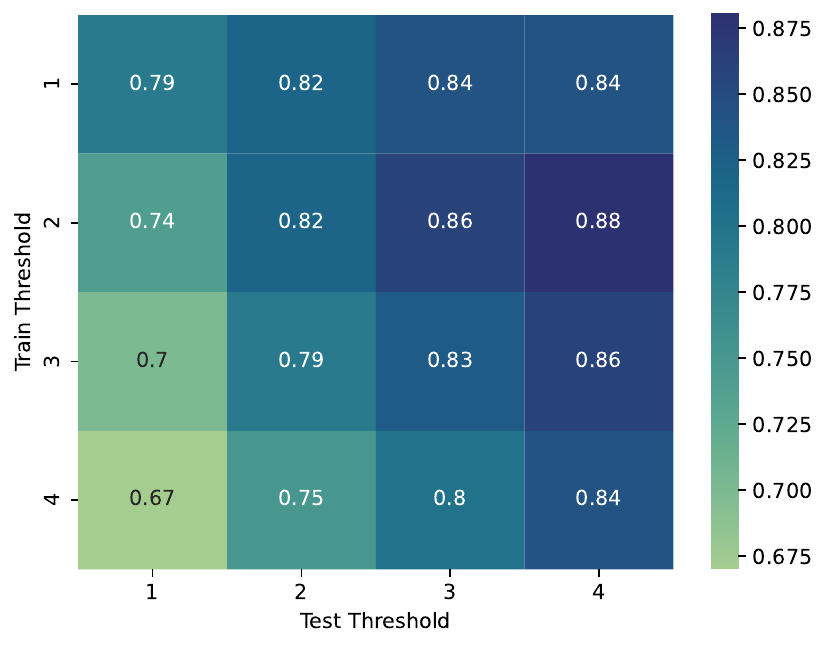}
        \caption{\label{pos-thresh} Macro average accuracy over each train and test pedestrian count threshold for radius $r=\SI{6}{\meter}$.}
     \end{figure}

\textbf{E2:}  When attempting to compare the performance across different radii in Fig.~\ref{all-res}, it is important to note that the test sets are not identical; although all audio content is identical, the labels and, therefore, class-proportions differ. 
The performance per class tends to be most balanced using radii 3 and \SI{6}{\meter}. The performance for radius \SI{1}{\meter} likely suffers due to pedestrian signals just outside the radius being labeled as no-activity, \pavan{while radii 6 and \SI{9}{\meter} see a slight decline} from the opposite issue: low-volume pedestrian signals on the edge are labeled as pedestrians while potentially not detectable from audio. 

\textbf{E3:} Figure~\ref{pos-thresh} visualizes the macro accuracy for each permutation of combinations of train threshold and test threshold for pedestrian count. We can make the following observations. First, as the threshold for the \textit{test} pedestrian count increases, a greater proportion of the samples are classified correctly. It is unsurprising that the classifier can perform better as an increased pedestrian count likely correlates to stronger signals and thus more easily detected frames. Second, performance generally decreases with increasing threshold for the \textit{training} pedestrian count, indicating that the classifier benefits from harder to classify training samples. 
In general, the performance seems to be best when trained with a low pedestrian count threshold and evaluated with a high pedestrian count threshold (upper right triangle).

\section{Conclusion}
\label{sec:conclusion}
    We have introduced the new large-scale dataset ASPED for the challenging task of detecting pedestrians from audio data. The dataset includes high quality audio recordings plus the video recordings used for labeling the data with pedestrians counts. 
    The baseline results indicate the feasibility of using audio sensors for pedestrian tracking, although the performance needs to be improved before systems become practically usable.
    
    Plans for future work include extending the dataset to additional environment such as locations with car traffic. Future directions of inquiry will include 
    \begin{inparaenum}[(i)]
        \item robustness against noise, 
        \item impact of the recording quality (sample rate etc.), 
        \item impact of weather such as wind on rain on the performance,
        \item accuracy of regression approaches to predict exact pedestrian counts, and 
        \item the performance of more sophisticated classification approaches for pedestrian detection. 
    \end{inparaenum}
    Accurate detection of pedestrians can offer valuable information and guidance for transportation infrastructure planning and urban planning in general. Knowing why people choose to walk in certain urban areas and the destinations they prefer can help improve pedestrian infrastructure, reduce overcrowding, and provide information about untapped opportunities for economic development. Future plans also include using the ASPED dataset for pedestrian route prediction using network flow models.

\vfill\pagebreak

\bibliographystyle{IEEEbib}
\bibliography{strings,uas,Subhro_Zoterobib}

\end{document}